\documentclass[aps,twocolumn,showkeys]{revtex4}
\pdfoutput=1
\usepackage{graphicx}                                       
\usepackage{amssymb}
\usepackage{wasysym}
\usepackage{caption}
\usepackage{lipsum}
\usepackage[a4paper,top=15mm,headheight=50mm,%
   bottom=15mm,left=12mm,right=12mm]{geometry}

\newcommand{\umafigura}[3]{
  \noindent%
  \begin{minipage}{\linewidth}% to keep image and caption on one page
  \makebox[\linewidth]{%        to center the image
    \includegraphics[width=\linewidth]{#1}}
   \captionof{figure}{\small #3}\label{#2}%      only if needed  
  \end{minipage}
}%fim-umafigura

\newcommand{\umatabela}[3]{ 
  \noindent%
  \begin{minipage}{\linewidth}% to keep image and caption on one page
  \makebox[\linewidth]{%        to center the image
    \includegraphics[width=\linewidth]{#1}}
   \captionof{table}{\small #3}\label{#2}% only if needed  
  \end{minipage}
}%fim-umafigura

\begin{document}

%____________________________________________________________________
%
%____________________________________________________________________
%
%____________________________________________________________________
%
%____________________________________________________________________
%                                RESUMO
\title{Results of DC Measuments on a Coaxial Resistor}
%____________________________________________________________________
%                                AUTORIA
\author{
  F. A. Silveira${}^{}$\footnote{Email address: {\it fsilveira@inmetro.gov.br}} 
%and
%  R. T. B. Vasconcellos${}^{}$\footnote{Email address: {\it rtbvasconcellos@inmetro.gov.br}}
}

\affiliation{
  %$^1$Instituto de F\'\i sica, Universidade Federal Fluminense,\\
  %Avenida Litor\^anea s/n, 24210-340 Niter\'oi RJ, Brazil
  %\vskip .3cm
  Instituto Nacional de Metrologia, Qualidade e Tecnologia -- Inmetro,\\
  Avenida N. S. das Gra\c cas 50, 25250-020 D. Caxias RJ, Brazil}

\date{\today}

\begin{abstract}

In this report, we briefly describe the most recent results from tests on the
calculable resistance prototypes from the Electrical Standards Metrology Laboratory
from Inmetro (Lampe/Inmetro). These prototypes were built using a different project from
that originally proposed by Haddad \cite{haddad}, in which the resistive element is mechanically 
fixed. A new fixation system of the resistive wire elements was conceived, in order to
overcome experimental difficulties, and the results of new measurement runs are presented here.

\end{abstract}


\maketitle

%____________________________________________________________________
%
%____________________________________________________________________
%
%____________________________________________________________________
%
%____________________________________________________________________
%                           INTRODUÇÃO
\section{Introduction}
\label{intro}

In order to relate capacitance standars to quantum Hall effect, we need an intermediary stage
on a quadrature bridge, where comparisons are made against a standard resistor
with calculable frequency dependence \cite{qhe}. Lampe has two protoypes of such 
resistors under development \cite{lampe1,lampe2,lampe3,lampe4}; the 
project is shown in Figures \ref{fig1a} and \ref{fig1b}.

\umafigura{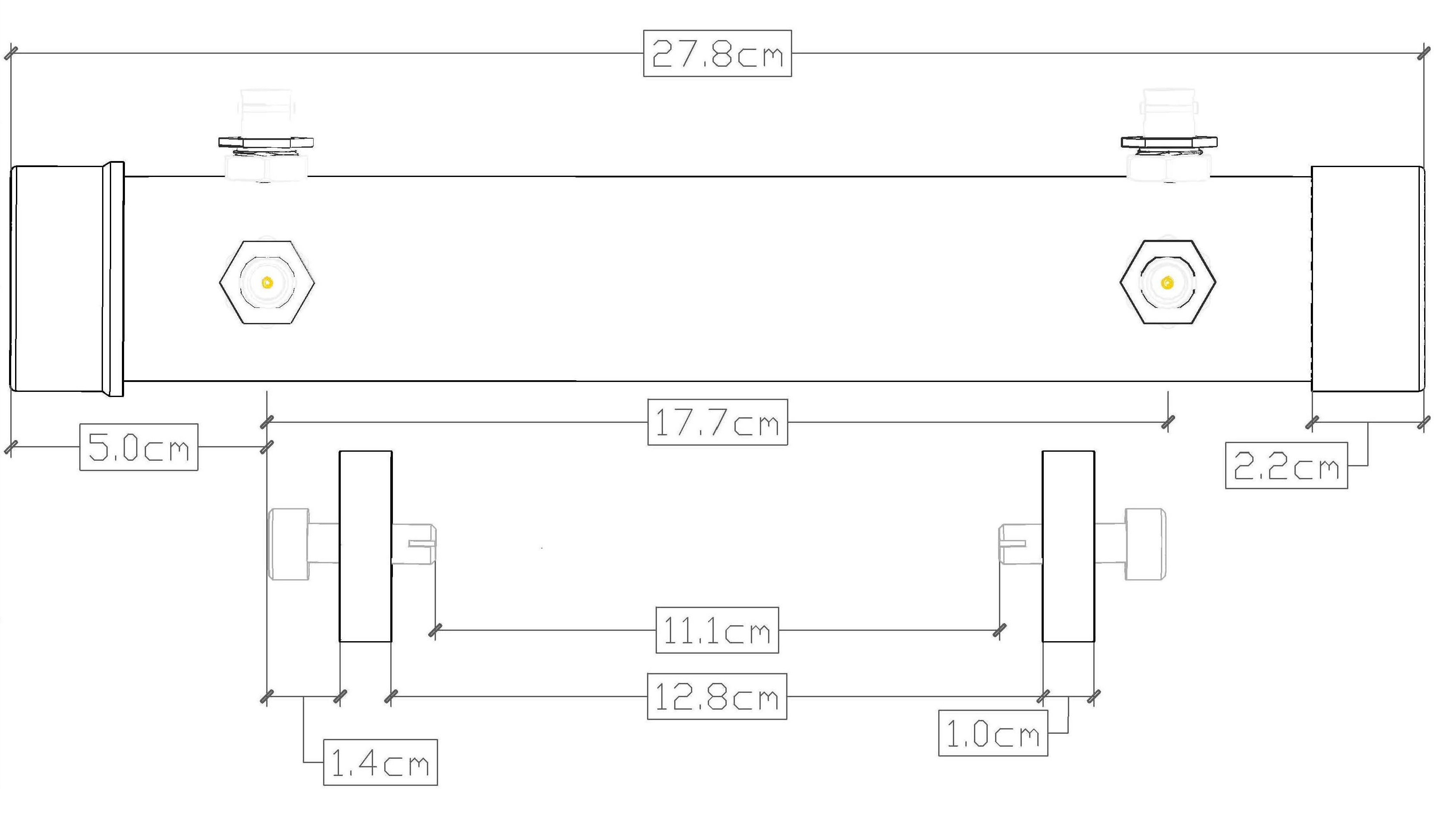}{fig1a}{Dimensional profile of Lampe's calculable
resistor prototypes \cite{lampe2}. Dimensions are consistent
with a 1k$\Omega$ resistor, when the resistive element is an EvanOhm
wire of 14 $\mu$m.}

\umafigura{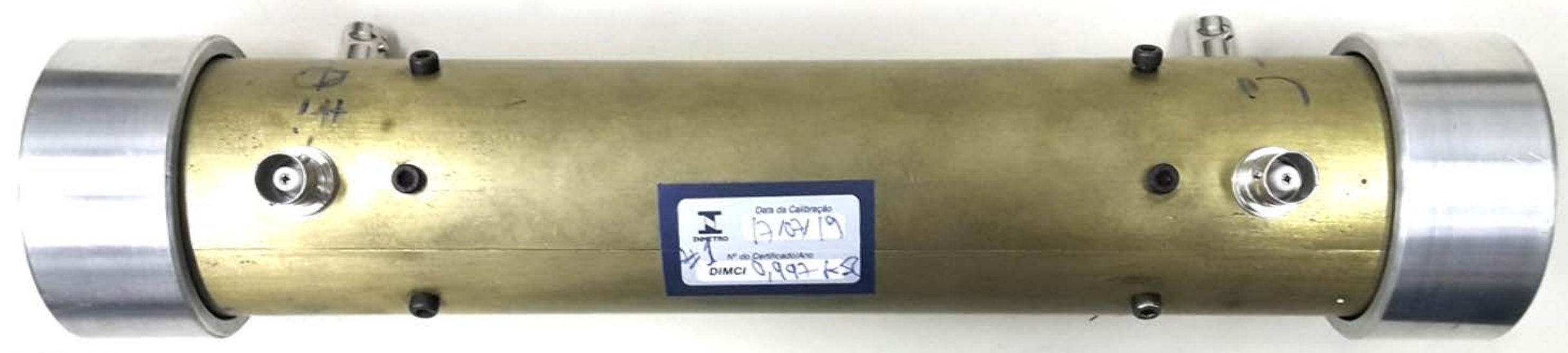}{fig1b}{Picture of one of the prototypes after assembling, 
by the beginning of 2018.}

We intend to test these prototypes before the design of a definitive model for
Lampe's calculable resistor, and an important such test
consists of DC measurements of the value of resistance, described in the
next section. Parameters and results of these measurements are presented in 
Sec. \ref{results}, and a few remarks are left to Sec. \ref{conclusao}.

%____________________________________________________________________
%
%____________________________________________________________________
%
%____________________________________________________________________
%
%____________________________________________________________________
%                   PRINCIPIOS DE FUNCIONAMENTO

\section{Wire Fixation and Measurements}
\label{system}

The resistors present a few construction problems,
due to some operational difficulties faced by Inmetro's precision
workshop by the time, and some of these problems
make the resistors unfit for
agreement to frequency dependency calculations.
But despite the construction flaws,
DC stability and temperature dependence precision measurements are possible,
and tests have been consistently carried since the beginning of 2018,
date of the first assembly of the prototypes.

The most recent DC measurement runs, carried between 07/17 and 08/10/2019,
had the purpose of testing a trial wire fixation system on the
support screws. It consists of grinding the end of the screws, creating two plane 
surfaces parallel to the screw axis,
and drilling a 4mm hole perpendicular to these planes.

\umafigura{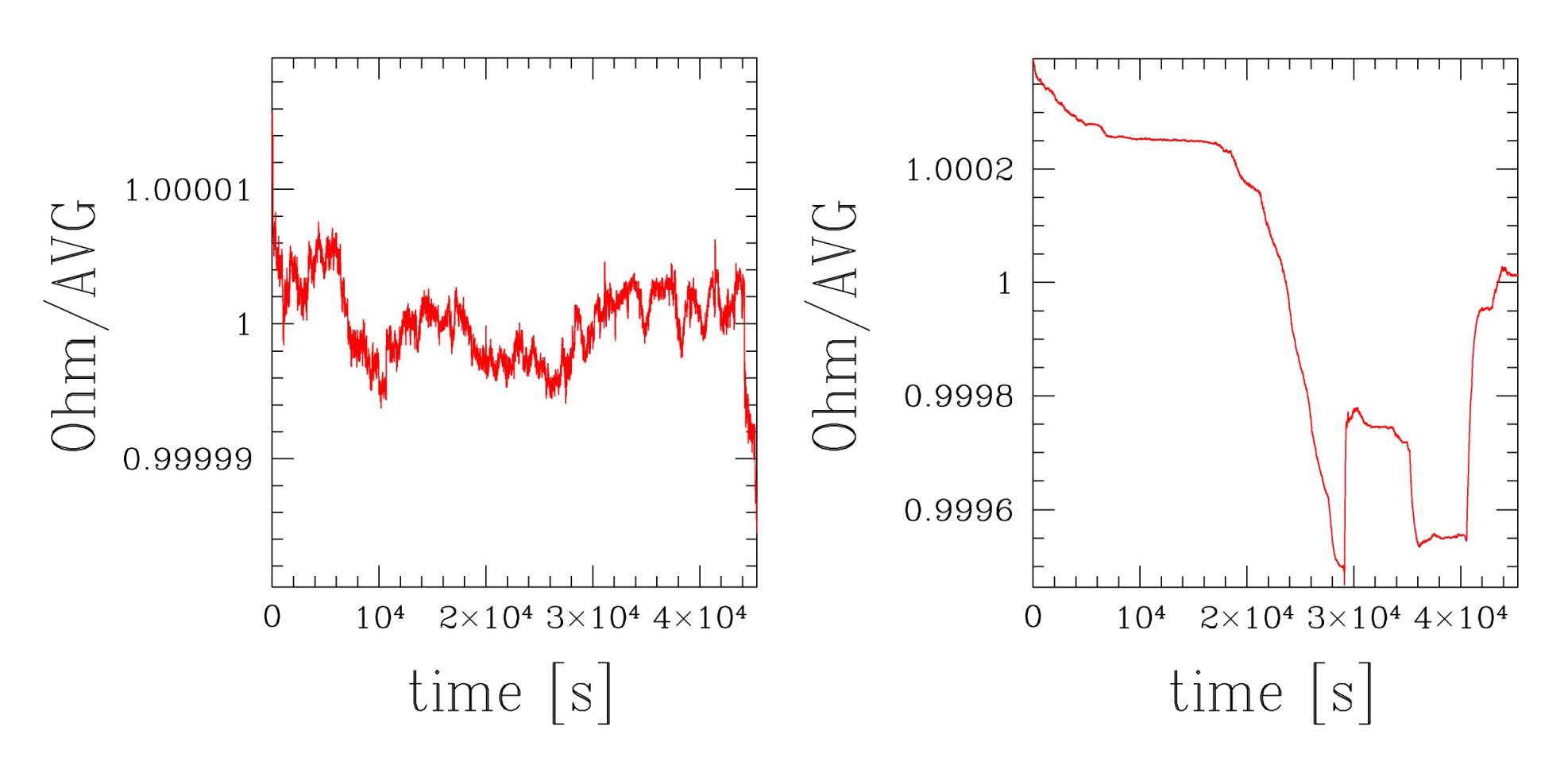}{fig2}{Relative resistance measurements taken between 07/17 and 07/18/2019
in A (left) and B. Interval coverage is approximately 12h. Prototype B
has erratic behaviour and suggests that the wire is still accomodating to its
support. The vertical axis of each figure shows the ratio between instant resistance
value and its average; horizontal axis shows time, in seconds.}

\umafigura{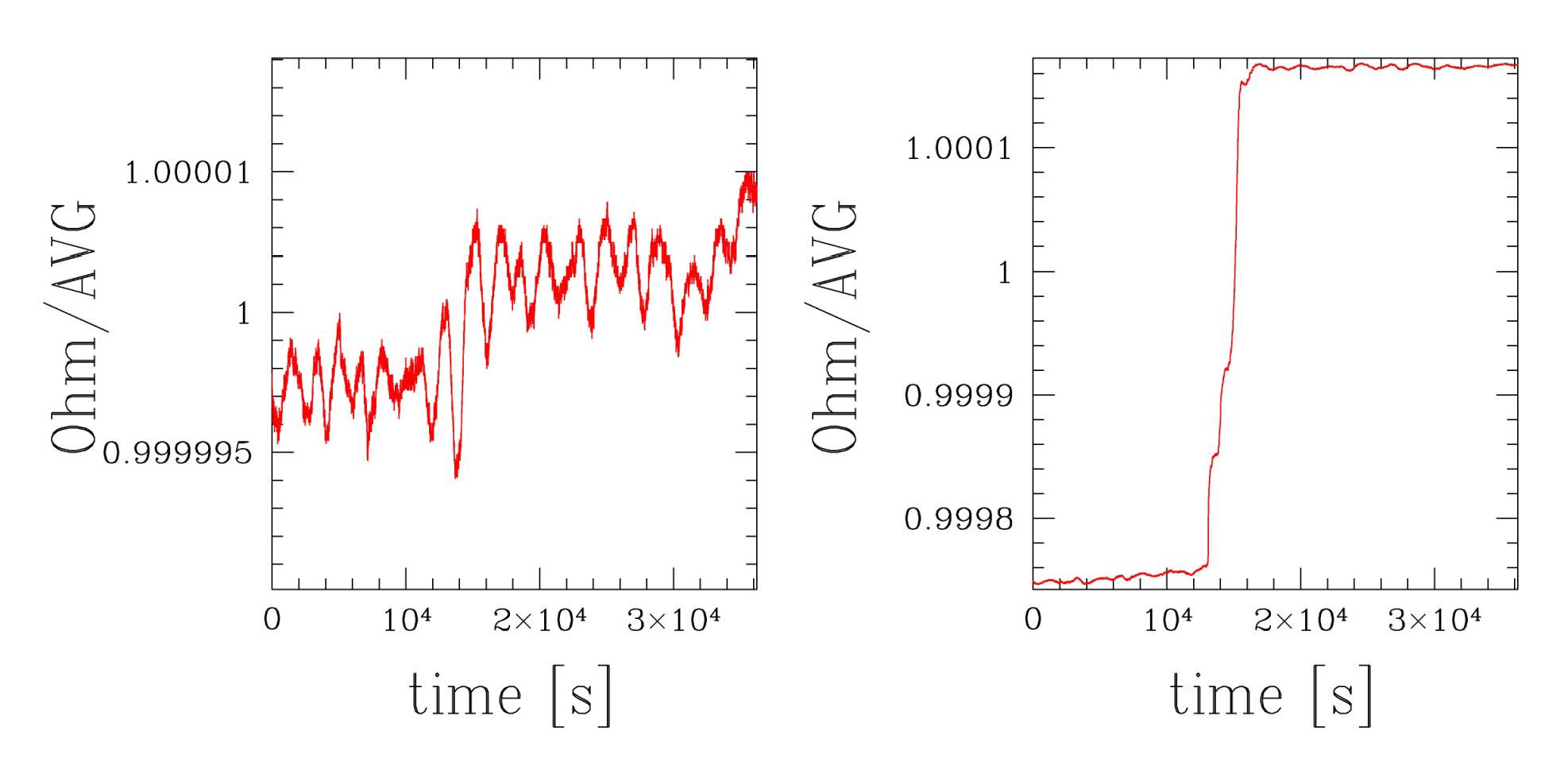}{fig3}{Relative resistance measurements taken between 07/29 and 07/30/2019
in A (left) and B. Interval coverage is approximately 10h. Protoype B persists in
acommodation. Prototype A begins to show periodic fluctuations correlated with 
environment temperature oscilations.}

At this stage, there has been no special care as to geometric aspects of the
grinding and drilling (central allignment of the hole or the direction of the planes,
for instance), as the main purpose was to test for mechanical stability of
the wire fixation only. Rough tools were employed in the grinding and
drilling of the screws ends. That choice was made in order to save the time of 
sending a new project to the precision workshop.

The wire is fixed by passing it through the hole and pressing it against the
grinded surfaces of the screws with a pair of metal sheets, sandwich-like, and then
tightening the whole set with a screw/nut pair of 3mm.

%____________________________________________________________________
%
%____________________________________________________________________
%
%____________________________________________________________________
%
%____________________________________________________________________
%                             RESULTS

\section{Results}
\label{results}

From this point on, the two prototypes will be identified just as A and B. 
At the assembling day, the nominal value of the resistors was determined by
a hand-held DMM, just for record keeping. Nominal value of A was determined to be
0,997 k$\Omega$; B's was 1,013 k$\Omega$. The tests, with measurement
runs between 07/17 and 08/10/2019, have the results summarized in Figures \ref{fig2} to \ref{fig4}.
Measurements were taken using two HP 3458A digital multimeters at a 4-terminal configuration, with an
interval of 5s between measurements and a aquisition window of 2s; identical
setup was used in \cite{lampe2,lampe3,lampe4}.

Prototype B shows, in Figs. \ref{fig2} and \ref{fig5}, a behaviour already observed  previously
\cite{lampe1}, under different schemes of fixation of the wire. The
resistance leaps of B are attributed to a slow accomodation of
the wire. In the course of these experiments, this accomodation period has taken about
15 days; after this period, the fluctuations on the measured resistance of the two resistors 
begin to approach each other (Fig. \ref{fig4}),
until differences between the two
prototypes become negligible, in practice (Fig. \ref{fig5}).

\umafigura{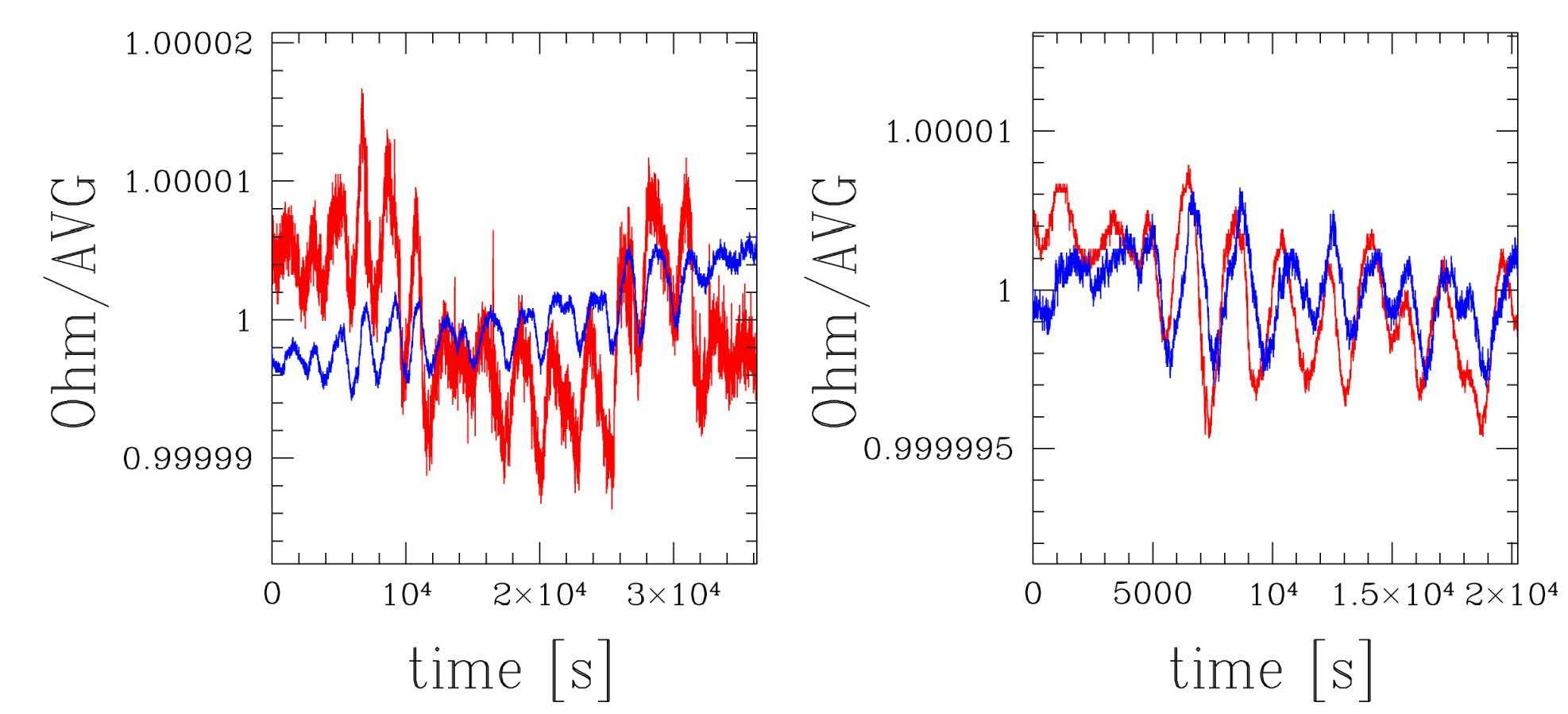}{fig4}{Relative resistance measurements taken between 08/01 and 08/10/2019
in A (red line) and B. Interval coverage is of 5 and 10h, respectively. Prototype B
has finished its accomodation period; both A and B pursue the periodic oscilations
of temperature that are caused by action of the environment control system of the laboratory.}

%____________________________________________________________________
%
%____________________________________________________________________
%
%____________________________________________________________________
%
%____________________________________________________________________
%                           CONCLUSION

\section{Final Remarks}
\label{conclusao}
Taking data from Fig. \ref{fig5}, the final characteristics of A and B can be summarized
in Table \ref{tab1}.
The drift values of resistors are still too high for a resistance standard, though after
accomodation period, both drift values and uncertainties became consistent between the two prototypes.
Measured resistance presents low uncertainties, but they're also about 10 times the expected 
values if the models are to work as calaculable standards.

We also still lack available micrometers to aid in the positioning of the wires in their nests.
Linear resistivity of the wire is around 9.7 k$\Omega/$m \cite{lampe1}, and 
its length should handled to 0.1 $\mu$m, if a 0.001 $\Omega$ precision in the setting of
resistors nominal values is
expected, which explains the large deviations of nominal 
values of A and B from 1 k$\Omega$.

More tests are in order, but we still seek for improvement on the support system of the wire
that complies to the model on \cite{lampe1,lampe2,lampe3,lampe4}, while
displays low costs and ease of mounting and assembling.

\umafigura{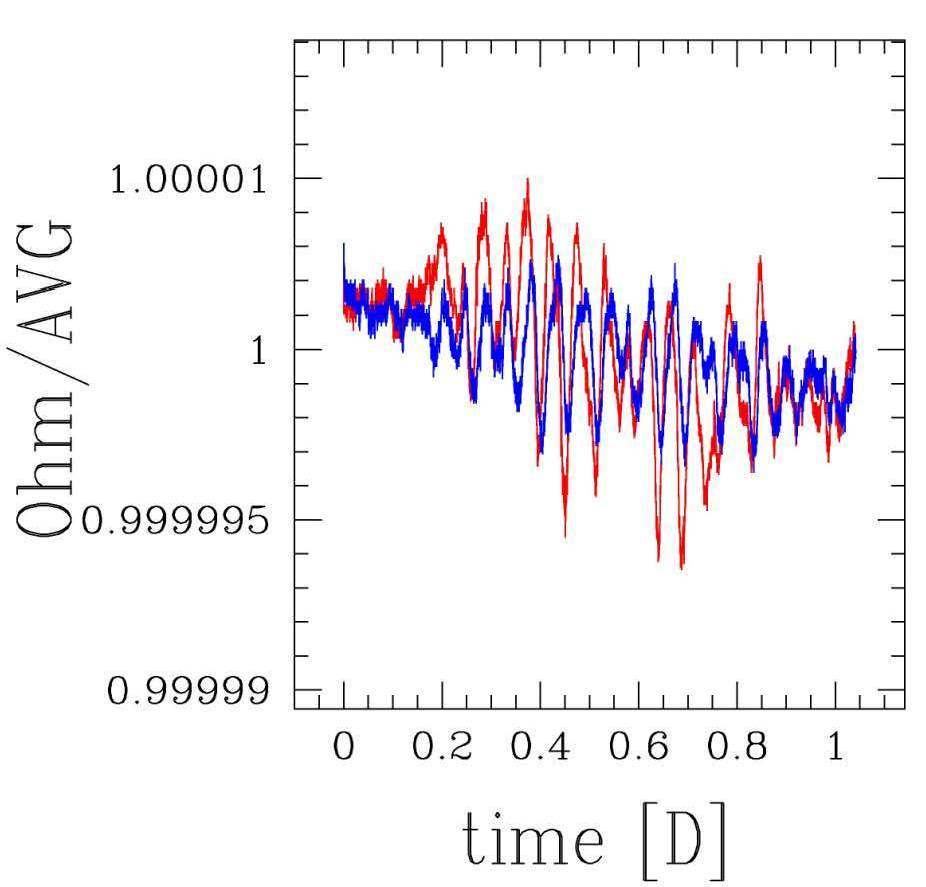}{fig5}{Relative resistance measurements taken in 08/07/2019
in A (red line) and B. Interval coverage is approximately 24h.
The vertical axis of each figure shows the ratio between instant resistance
value and its average; horizontal axis shows time, in days.}
\umatabela{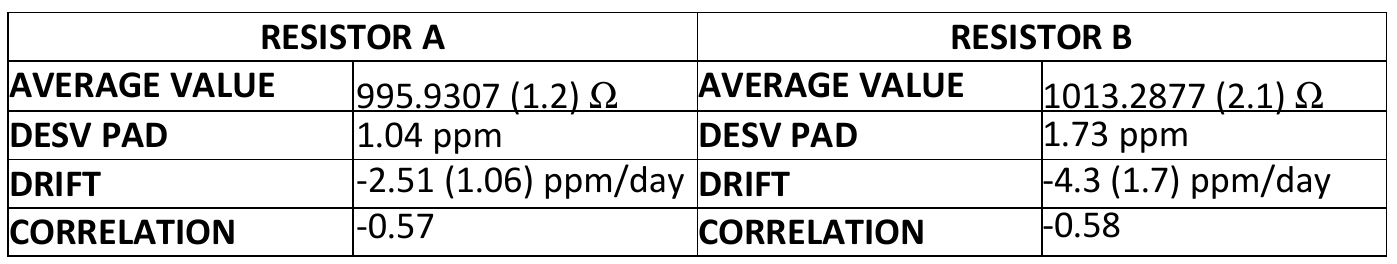}{tab1}{Summary of A and B properties after DC stabilization.}

\ \vfill

%____________________________________________________________________
%
%____________________________________________________________________
%
%____________________________________________________________________
%
%____________________________________________________________________
%                                BIBLIOGRAFIA

%Gives vertical space between 'Conclusions' and
%'References' sections when in twocolumn mode!
%\ \onecolumngrid\ \vfill\twocolumngrid

%____________________________________________________________________
%                                   FIM
\end{document}